\documentclass[preprint,12pt]{elsarticle}

\usepackage{amssymb}

\usepackage{amsmath,amssymb,amsfonts}
\usepackage{algorithmic}
\usepackage{algorithm}
\usepackage{graphicx}
\usepackage{textcomp}
\usepackage{multirow}
\usepackage{multicol}
\usepackage{pbox}
\usepackage{pifont}

\usepackage{titlesec}
\titlespacing*{\section}{0pt}{*1}{*1}
\titlespacing{\subsection}{0pt}{*1}{*1}
\titleformat{\subsubsection}[runin]{\itshape}{\thesubsubsection)}{1em}{}
\titlespacing*{\subsubsection}{\parindent}{0pt}{*1}

\def\x{{\mathbf x}}

\newcommand\figref{Fig.~\ref}
\newcommand\tabref{Table~\ref}
\newcommand\secref{Section~\ref}
\newcommand{\mycomment}[1]{}
\newcommand{\cmark}{\ding{51}}%
\newcommand{\xmark}{\ding{55}}%

\begin{document}

\begin{frontmatter}



\title{Multi-view Cardiac Image Segmentation via Trans-Dimensional Priors}


\affiliation[inst1]{School of Electronic Engineering and Computer Science, Queen Mary University of London, UK}
\affiliation[inst2]{Queen Mary's Digital Environment Research Institute (DERI), London, UK}
\affiliation[inst3]{School of Biomedical Engineering and Imaging Sciences, King’s College London, UK}
\affiliation[inst4]{Department of Computer Science, University College London, UK}
\affiliation[inst5]{School of Engineering and Materials Science, Queen Mary University of London, UK}

\author[inst1,inst2]{Abbas Khan}
\author[inst2,inst3]{Muhammad Asad}
\author[inst4]{Martin Benning}
\author[inst2,inst5]{Caroline Roney}
\author[inst1,inst2]{Gregory Slabaugh}

\begin{abstract}
We propose a novel multi-stage trans-dimensional architecture for multi-view cardiac image segmentation. Our method exploits the relationship between long-axis (2D) and short-axis (3D) magnetic resonance (MR) images to perform a sequential 3D-to-2D-to-3D segmentation, segmenting the long-axis and short-axis images. In the first stage, 3D segmentation is performed using the short-axis image, and the prediction is transformed to the long-axis view and used as a segmentation prior in the next stage. In the second step, the heart region is localized and cropped around the segmentation prior using a Heart Localization and Cropping (HLC) module, focusing the subsequent model on the heart region of the image, where a 2D segmentation is performed. Similarly, we transform the long-axis prediction to the short-axis view, localize and crop the heart region and again perform a 3D segmentation to refine the initial short-axis segmentation. We evaluate our proposed method on the Multi-Disease, Multi-View \& Multi-Center Right Ventricular Segmentation in Cardiac MRI (M\&Ms-2) dataset, where our method outperforms state-of-the-art methods in segmenting cardiac regions of interest in both short-axis and long-axis images. The pre-trained models, source code, and implementation details will be publicly available.
\end{abstract}

\begin{graphicalabstract}
\centering
\includegraphics[scale=0.60]
{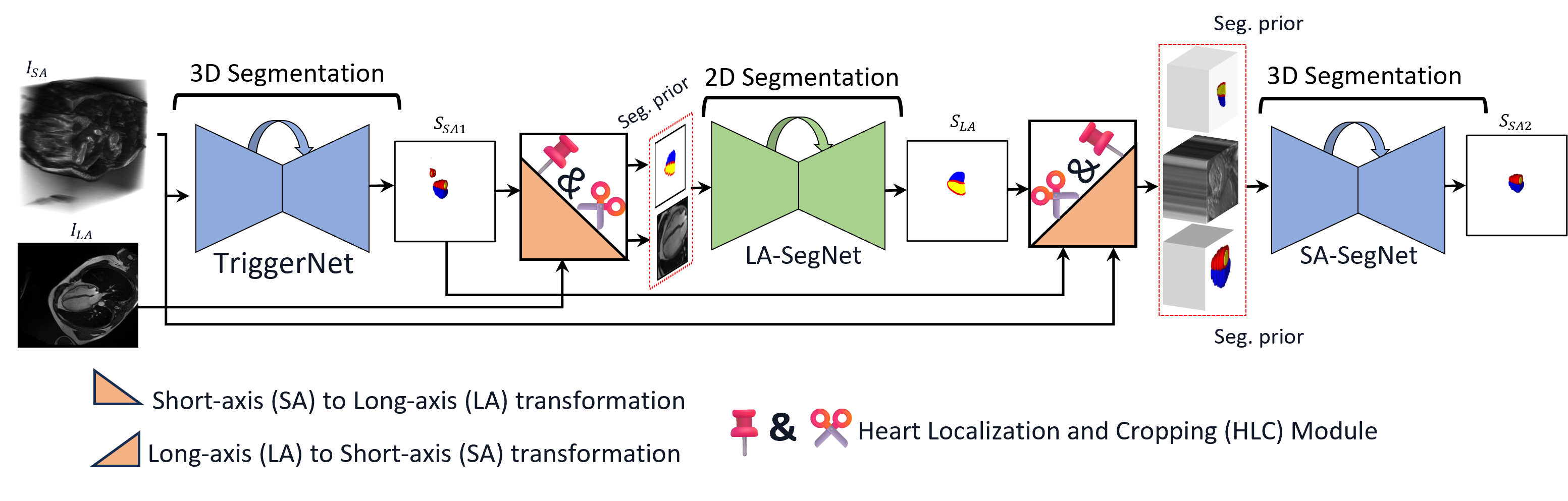}
\end{graphicalabstract}

\begin{highlights}
    \item We propose a sequential 3D-to-2D-to-3D approach for multi-view cardiac image segmentation by effectively utilizing the trans-dimensional segmentation priors (TDSP), which transform a segmentation from one view into another and serve as guidance. 
    \item The TDSP provides a robust anatomical reference at the network's input and encourages the network to produce anatomically plausible segmentation maps.
    \item We also introduce a Heart Localization and Cropping (HLC) module to focus the segmentation on the heart region only. This strategy reduces the computation for the second and third-stage segmentation network and eliminates false positive predictions.
    \item Extensive experiments are conducted to showcase the efficacy of the proposed pipeline utilizing the HLC module and TDSP, where our proposed method outperforms the state-of-the-art as well as methods on the M\&Ms-2 challenge leaderboard. 
\end{highlights}
    

\end{frontmatter}


\section{Introduction}
\label{sec:introduction}
Cardiovascular disease is the leading cause of death, with a yearly toll of 23.6 million lives due to heart disease and stroke globally \cite{greenfield2019cardiovascular}. This underscores the need to identify and treat cardiac disorders. Cardiologists have focused on early diagnosis as part of the clinical workflow \cite{stats}. Deep learning architectures have achieved a wide range of competencies for computational cardiac imaging \cite{moradi2023recent},\cite{kadem2022hemodynamic},\cite{fotaki2022artificial} including segmentation \cite{chen2020deep},\cite{li2023multi},\cite{li2022towards}. 

Modern non-invasive medical imaging techniques, such as ultrasound, magnetic resonance imaging (MRI), and computed tomography, are widely used to capture detailed images of the structure and function of the heart and its associated vessels \cite{rajiah2023cardiac}. However, detection of disease and quantification often requires a laborious process of manual segmentation to identify the areas of the anatomy in these scans. Recent advances in artificial intelligence \cite{chen2023artificial} are improving automation to segment a medical image into meaningful areas of interest. In the context of cardiac imaging, areas of interest include the left atrium, right atrium, left ventricle, right ventricle, and myocardium to diagnose different cardiac pathologies. Many image segmentation methods have been proposed, including active shape models \cite{de2003adapting}, active appearance models \cite{ mitchell20023}, atlas-based methods \cite{kiricsli2010evaluation}, convolutional neural network (CNN)-based approaches \cite{baccouch2023comparative},\cite{xu2023deep} including those with self-attention-based architectures \cite{huang2022missformer},\cite{deng2021transbridge}.  

Among the successful cardiac image segmentation methods, most rely on a single view, i.e., short-axis (SA) or long-axis (LA), where the segmentation is performed. However, capturing both SA and LA MR images is considered standard practice \cite{kramer2020standardized}, \cite{petersen2015uk}, and segmentation of one view can be utilized to improve the segmentation of the other. Here, we propose a novel framework that performs accurate cardiac image segmentation by transferring the segmentation of one view to guide the segmentation of the other. Our proposed method sequentially utilizes the multi-view images. Despite being based on single encoder-decoder segmentation networks, the proposed pipeline still benefits from multi-view data.

\figref{fig1}(d) depicts the overall architecture of the proposed pipeline. TriggerNet, which functions as a 3D segmentation model, generates the initial segmentation for the short-axis denoted as $S_{SA1}$.
Subsequently, utilizing the transformation parameters for the given volume, $S_{SA1}$ undergoes transformation to the LA view to produce a SA-to-LA map (SA2LAmap), a trans-dimensional segmentation prior. The SA2LAmap and LA image ($I_{LA}$) are fed as input to the Heart Localization and Cropping (HLC) module, resulting in a cropped $I_{LA}$ and SA2LAmap containing only the heart region.  The SA2LAmap is input along with the cropped $I_{LA}$ to the LA-SegNet model that generates a segmentation for the long-axis named $S_{LA}$.

We next refine the short axis segmentation $S_{SA1}$. $S_{LA}$ is transformed to the SA view, resulting in the LA-to-SA map (LA2SAmap), another trans-dimensional segmentation prior. Here, we again use the HLC module to obtain cropped LA2SAmap, SA image ($I_{SA}$), and the TriggerNet output ($S_{SA1}$). Finally, the SA-SegNet utilizes these cropped outputs of the HLC module and generates the final segmentation for the short-axis named $S_{SA2}$.

In our proposed framework, integrating the segmentation from alternate views (SA to LA and LA to SA) acts as a segmentation prior, and provides HLC module guidance to remove the surrounding background regions and improve overall segmentation accuracy for the respective views. This framework enables LA-SegNet and SA-SegNet segmentation to outperform the existing state-of-the-art methods on the Multi-Disease, Multi-View \& Multi-Center Right Ventricular Segmentation in Cardiac MRI (M\&Ms-2) dataset's challenge \cite{martin2023deep}. The proposed framework efficiently utilizes the multi-view aspect of the M\&Ms-2 dataset, as the challenge provides the images and labels of both views (LA and SA) for each instance, compared to the previous datasets M\&Ms \cite{campello2021multi} and ACDC  \cite{bernard2018deep}.

\section{Related Work}
This section lists some of the most well-known deep learning-based segmentation architectures, including those unifying the power of CNN and self-attention-based mechanisms. We also detail existing cardiac image segmentation approaches, specifically from the M\&Ms-2 dataset's challenge \cite{martin2023deep} leaderboard and subsequent publications leveraging the dataset. 
We note that our proposed pipeline can be implemented with any segmentation backbone, provided that the architecture can segment both 2D and 3D views.

UNet \cite{ronneberger2015u} revolutionized deep learning-based medical image segmentation by proposing a symmetric encoder-decoder architecture. The encoder part extracts the features from the image, and the decoder reconstructs the segmentation map, while the skip-connections help to propagate information across different stages.  The no-new-Net (nnUNet) \cite{isensee2021nnu} is built on UNet and proposed an automatically configurable segmentation architecture. It can configure data pre-processing, network design, and post-processing for many medical image segmentation datasets. An overview is provided in \secref{sec:nnunet}. ResUNet \cite{diakogiannis2020resunet} is an encoder-decoder architecture based on the UNet model and also incorporates knowledge of residual connections \cite{he2016identity}, atrous convolutions \cite{chen2017deeplab}, and pyramid scene parsing (PSP) pooling \cite{zhao2017pyramid}. Each convolution block is replaced with a residual block to achieve consistent training with the increased network depth, atrous convolutions help increase the receptive field, and PSP pooling enhances the network's performance by including background context information.

Inspired by the emergence of vision transformers \cite{dosovitskiy2020image} in computer vision regimes \cite{carion2020end}, many hybrid architectures that utilize multi-head self-attention (MHSA) \cite{vaswani2017attention} have been proposed for medical image segmentation. TransUNet \cite{chen2021transunet} is a UNet architecture that utilizes both CNN and self-attention. This includes a transformer-based encoder that extracts features from images and a CNN-based decoder that upsamples the encoded features. UTNet \cite{gao2021utnet} is also a hybrid architecture integrating transformer and CNN for medical image segmentation. It proposes a revised MHSA mechanism to reduce the complexity of the model.  In addition, a hybrid layer utilizing CNN and revised MHSA is incorporated into the encoder and decoder stages. 
The Multi-Compound Transformer (MCTrans) \cite{ji2021multi} aims to combine rich features and semantic structures into multi-scale convolutional features using self-attention. The MCTrans transforms convolutional features as a sequence of tokens to perform intra- and inter-scale self-attention across multiple scales.
A multi-view and transformer-based architecture named Transfusion was proposed by \cite{liu2022transfusion} to correlate and fuse data coming from SA/LA views. It proposed Divergent Fusion Attention (DiFA), which combines features from different views using multi-scale self-attention.
Al Khalil et al. \cite{al2023reducing} proposed a three-stage approach: firstly, the region of the heart is detected using a regression model; secondly, a GAN-based augmentation technique is used for image synthesis to increase the diversity of the training data for segmentation tasks. More specifically, their approach generates more examples of pathologies to balance instances of pathological and normal cases. Lastly, the late-fusion segmentation approach combined with intensity transformations is utilized to generate the final segmentation map.

Sun et al. \cite{sun2022right} utilized labels from the end-diastolic and end-systolic phases through an intensity-based image registration approach. These registered labels increase the size of the training set. Arega et al. \cite{arega2021using} relied on the MRI-specific based, intensity, and spatial data augmentation techniques to improve the generalization and robustness of their segmentation models.  
In \cite{jabbar2021multi}, a multi-view SA-LA Network was proposed to simultaneously segment the RV blood pools in both the SA and LA views. It merged the bottleneck features from both the SA and LA and combined the labels of the left ventricle (LV) and myocardium (MYO) to generate a label that aids with contextual information to better segment the right ventricle (RV). Another multi-encoder-decoder network (xUnet) is proposed by \cite{queiros2021right} to simultaneously process the SA and LA views. It utilizes a pre-processing step where both views are centered and rotated to match their axes. A spatial transformer multi-pass feature pyramid (Tempera) \cite{galazis2021tempera} segments the RV in both SA and LA cardiac MR images. Tempera is based on the multi-scale feature pyramid network from \cite{lin2017feature} and transforms the SA features to LA via a geometric target spatial transformer. InfoTrans \cite{li2021right} proposed a nnUNet-based architecture, where the first 2D-nnUNet segments the LA views and then utilizes the LA prediction to crop the region of interest (ROI) from SA views.
The Refined Deep Layer Aggregation (RDLA) \cite{liu2021refined} proposed a two-stage 2D architecture, using DLA-34 stride-2 network \cite{yu2018deep} as the backbone. The LA and SA images are segmented independently, followed by a refinement step by utilizing the complementary information of another view along with the images.

\begin{figure}[t!]
\centering
\includegraphics[scale=0.48,trim={2.0em, 0.0em, .5em, 4.5em}]{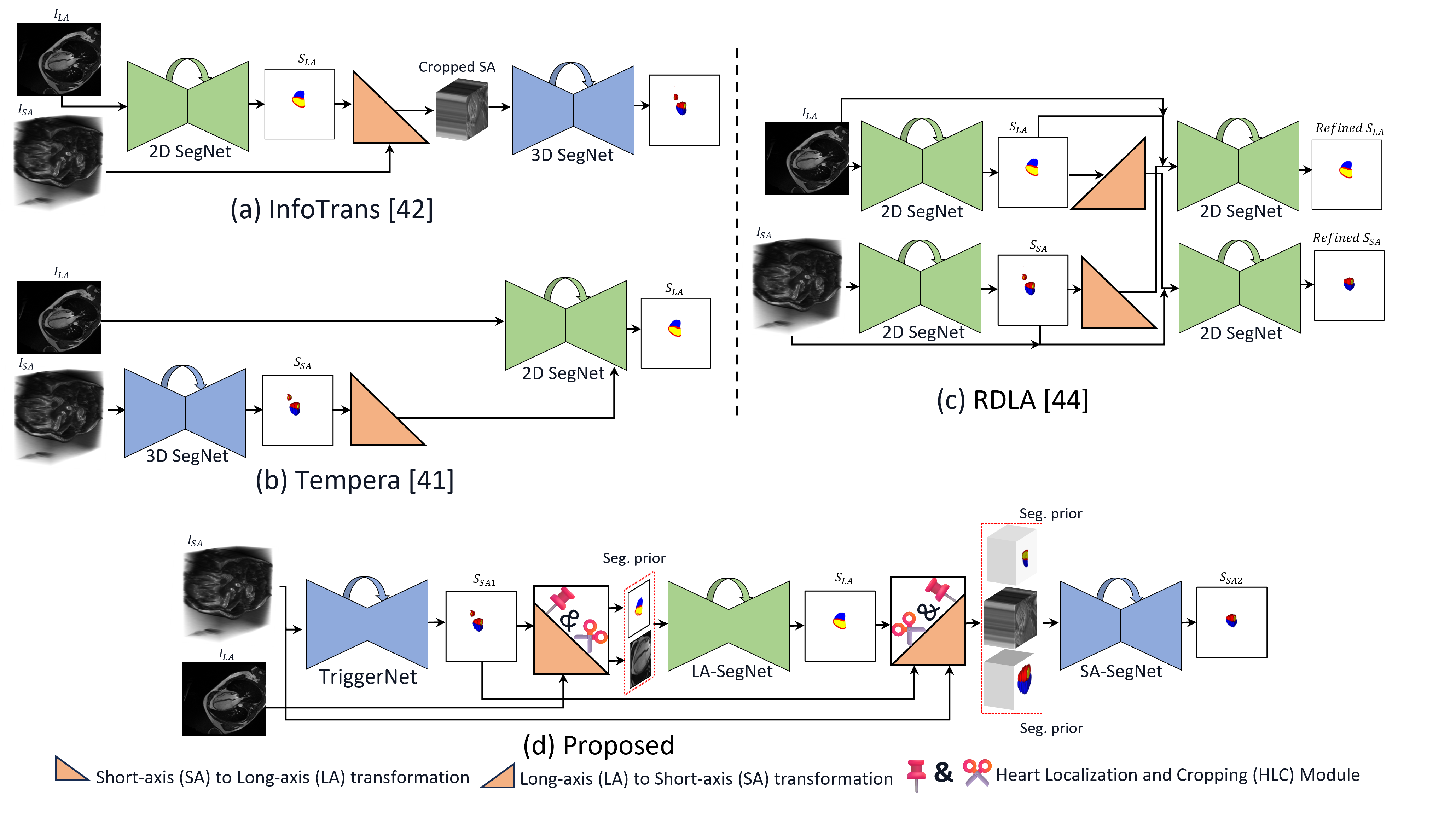}
\caption{Overview of other related methods and our proposed pipeline. The proposed framework (d) segments $I_{SA}$ using TriggerNet to produce $S_{SA1}$. It then generates $S_{LA}$ using LA-SegNet and refines $S_{SA2}$ using SA-SegNet,  along with relevant segmentation prior and image. Compared to (a), (b), and (c), our method (d) performs a two-way transformation (LA2SA and SA2LA) along with the utilisation of transformed maps as guidance for the HLC module.}
\label{fig1}
\end{figure}

Our proposed approach unites the strengths of \cite{galazis2021tempera}, \cite{lin2017feature}, and  \cite{liu2021refined}, as shown \figref{fig1}  and introduces a sequential approach in which each network benefits from the previous one, without introducing additional parameters compared to other multi-stage approaches such as \cite{al2023reducing}. 
InfoTrans \cite{lin2017feature} performs information transition only from an LA to an SA network and utilizes the transformed SA map (LA2SAmap) only to extract the region of interest (ROI) from original SA images. Compared to this method, we introduce the transformation from LA to SA to obtain the LA2SAmap and from SA to LA to obtain the SA2LAmap. Additionally, we also take advantage of these transformation maps and utilize them as segmentation priors for anatomically plausible predictions.
The Tempera \cite{galazis2021tempera} architecture only transforms SA to LA and segments each view twice using a two-stage methodology. Compared to Tempera, the proposed pipeline performs a complete two-way transformation, using the SA2LAmap and LA2SAmap to achieve 3D-to-2D-to-3D segmentation. Furthermore, Tempera does not utilize the transformed map to localize and crop the heart region in intensity images, as proposed by the HLC module. Hence, both stages in Tempera utilize full-scale images. 
Compared to \cite{liu2021refined}, which is implemented using four 2D networks, the proposed refinement strategy uses three networks, where we use a 3D network for SA images to effectively exploit the 3D spatial context and a single 2D network can achieve state-of-the-art performance using the transformation from TriggerNet as a segmentation prior. In addition, in RDLA architectures, the second-stage networks have the same complexity as the first-stage networks without taking advantage of alignment and first-stage predictions. However, in the proposed setting, the second-stage networks, i.e., LA-SegNet and SA-SegNet, receive images with a lower in-plane spatial resolution (images containing only heart regions) and have fewer encoder-decoder stages, which results in reduced computational complexity, as shown in \figref{param} (further details in \secref{ablation}). Al Khalil et al. \cite{al2023reducing}  employed a regression-based neural network as a crucial component for heart region detection within SA and LA images. While practical, this regression-based neural network brings additional trainable parameters, thereby increasing the complexity of the overall pipeline. In contrast, our proposed HLC module achieves the exact heart region localization and cropping task as their regression-based model without introducing additional trainable parameters. Instead, the HLC module utilizes segmentation maps from the previous network and crops the heart regions, thus offering a streamlined alternative to the regression-based approach. Our contributions can be summarized as follows:
\begin{enumerate}
    \item We propose a sequential 3D-to-2D-to-3D approach for multi-view cardiac image segmentation by effectively utilizing the trans-dimensional segmentation priors (TDSP), which transform a segmentation from one view into another and serve as guidance. 
    \item The TDSP provides a robust anatomical reference at the network's input and encourages the network to produce anatomically plausible segmentation maps.
    \item We leverage the TDSP and introduce a Heart Localization and Cropping (HLC) module to focus the segmentation on the heart region only. This strategy reduces the computation for the second and third-stage segmentation network and eliminates false positive predictions.
    \item Extensive experiments are conducted to showcase the efficacy of the proposed pipeline utilizing the HLC module and TDSP, where our proposed method outperforms the state-of-the-art as well as methods on the M\&Ms-2 challenge leaderboard. 
\end{enumerate}
\section{Proposed Framework}
\figref{fig1}(d) depicts our proposed framework, where the pipeline starts with trigger network (TriggerNet), followed by the transformation of its predictions $S_{SA1}$ to the LA view, resulting in SA2LAMap. As a pre-processing step for SA to LA transformation, the header information of the original SA image ($I_{SA}$) is applied to $S_{SA1}$ to ensure the matching of all properties of the $I_{SA}$ and $S_{SA1}$.
The SA2LAMap is used to remove the unrelated non-cardiac areas of the original LA image ($I_{LA}$) utilizing the HLC module and as input to the LA-SegNet through concatenation with the cropped $I_{LA}$ as a segmentation prior. The output from LA-SegNet is restored to its original size, followed by copying all the metadata information from the $I_{LA}$ to preserve the header information of the $I_{LA}$ in $S_{LA}$.
The final segmentation for LA ($S_{LA}$) is transformed to the SA view using the $I_{SA}$ to obtain LA2SAMap.
Following the same process, the HLC module utilizes the LA2SAMap and $I_{SA}$ to localize and crop the heart in a full-scale image.
Here, we also cropped and concatenated the $S_{SA1}$ from the TriggerNet (further details are provided in the ablation in \secref{ablation}). Finally, the $S_{SA2}$ is restored to its original size.

All three networks (TriggerNet, LA-SegNet, and SA-SegNet) are trained independently. The downstream tasks, such as LA-2-SA transformations and vice versa, and the HLC module are applied when the previous network outputs are available. However, at inference, the full framework is used sequentially to perform 3D-to-2D-to-3D segmentation output results for both the LA and SA images.

The following subsections will list the details of each step, segmentation networks, HLC module, and transformation process in the proposed pipeline shown in \figref{fig1}.
\subsection{Segmentation Networks}
\label{sec:nnunet}
Segmentation networks used within the proposed framework, i.e., TriggerNet, LA-SegNet, and SA-SegNet, are implemented using nnUNet \cite{isensee2021nnu}. The nnUNet is built upon the original U-Net architecture with modifications and improvements adopted for medical image segmentation tasks. It stands out from UNet due to its ability to configure the architecture and hyperparameters automatically during training. Moreover, nnUNet achieved the best results in the challenge cohort of  M\&Ms-2 \cite{martin2023deep}, and the proposed pipeline is built using its architecture as a baseline. The nnUNet architecture has three major components: (i) Encoder, (ii)  Decoder, and (iii) Skip-Connections.

\begin{figure}[b!]
\centering
\includegraphics[scale=0.5]{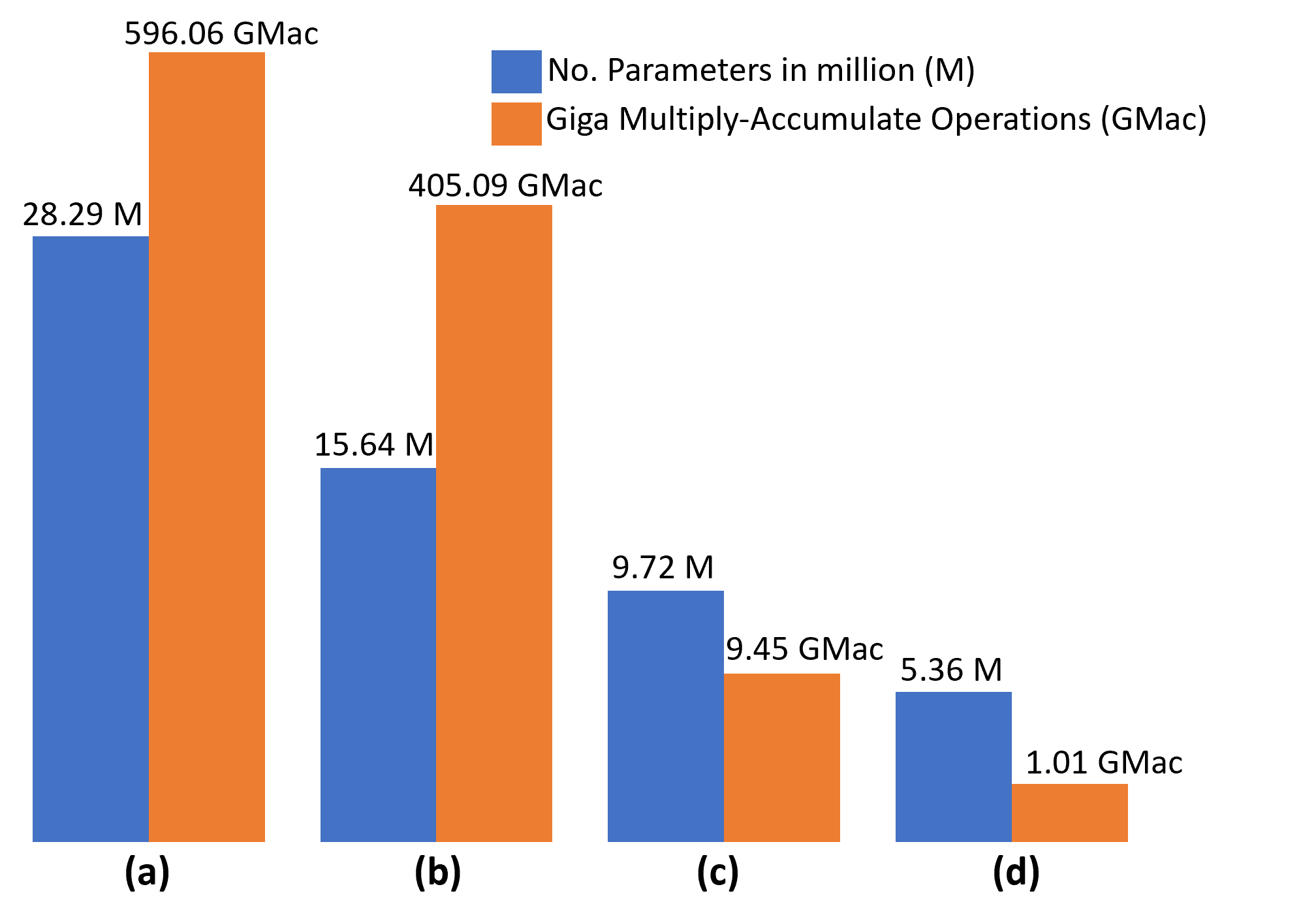}
\caption{Comparison of segmentation network complexities regarding the number of parameters and Multiply-Accumulate (MAC) operations with and without using the HLC module. (a) SA-SegNet without HLC Module (TriggerNet), and (Depth, Height, Width = 64,192,192), (b) SA-SegNet With HLC Module (SA-SegNet) and  (Depth, Height, Width = 112,128,112), (c) LA-SegNet without HLC Module and (Height, Width = 384,384), and (d) LA-SegNet with HLC Module and (Height, Width = 128,128).
 }
\label{param}
\end{figure}

The encoder extracts features from the input data by gradually increasing the number of features while reducing the spatial dimensions as it goes deeper. Each encoder block consists of two consecutive convolutions with a kernel size 3 (3$\times$3 for 2D/LA, and 3$\times$3$\times$3 for 3D/SA network). To reduce the spatial resolution, the features are again convolved with a kernel size of 3 and stride of 2. Each convolutional layer is followed by LeakyReLU activation and instance normalization.

%

The decoder reconstructs the segmentation map by progressively increasing the spatial dimension and reducing the number of features from the bottleneck layer. Each decoder block has two consecutive convolutions with a kernel size of 3, followed by a transpose convolution layer with a kernel size of 2 and stride 2. Similar to the encoder, each convolution layer is followed by LeakyReLU activation and instance normalization. The final convolution layer utilizes a sigmoid activation function with four kernels of size 1, where each kernel generates segmentation output for four classes, i.e., MYO, LV, RV, and background.

The skip-connections have been shown to improve segmentation methods \cite{drozdzal2016importance} for medical image segmentation tasks, and hence, we utilize skip connections in our proposed architecture. These skip-connections copy and concatenate the features from the contracting path from the encoder to the expanding path in the decoder for a better gradient flow during backpropagation and to recover the lost spatial information.

The number of encoder-decoder blocks can be different for each segmentation network shown in \figref{fig1}(d), and the corresponding computational complexity in \figref{param}, depending upon the spatial dimension of the input data. The TriggerNet gets spatial dimension images of 64 $\times$ 192 $\times$ 192, and it has five downsampling (encoder's block) and upsampling blocks (decoder's blocks). The LA-SegNet network gets the cropped input of spatial size 128 $\times$ 128 and has the four encoder and corresponding decoder blocks. The spatial dimension of input data for SA-SegNet is 112 $\times$ 128 $\times$ 112, and it also has four blocks for downsampling and upsampling. We also trained the LA-SegNet network on LA images without utilizing the HLC module for the ablation studies mentioned in \secref{ablation}. For this network, the nnUNet configures six encoder-decoder stages due to its large spatial dimension of 384 $\times$ 384.

\subsection{Heart Localization and Cropping (HLC) Module}
The foreground-background imbalance of pixels has been a fundamental issue for accurately segmenting medical images \cite{braytee2022comparative}. The foreground pixels occupy a smaller proportion of the image than the background objects.
It can significantly degrade the segmentation performance by forcing the model to focus more on the background pixels due to their majority compared to the foreground pixels \cite{yudistira2020prediction}. A common way of solving this issue is by designing the loss functions for segmentation, which can weigh the foreground pixels more than the background pixels \cite{salehi2017tversky}, \cite{lin2017focal}. However, in the proposed framework, this problem is solved intrinsically to some extent, as we cropped the original full-scale image using the segmentation prior, resulting in a reduction of background search space.
\begin{figure}[t!]
\centering
\includegraphics[scale=0.8]{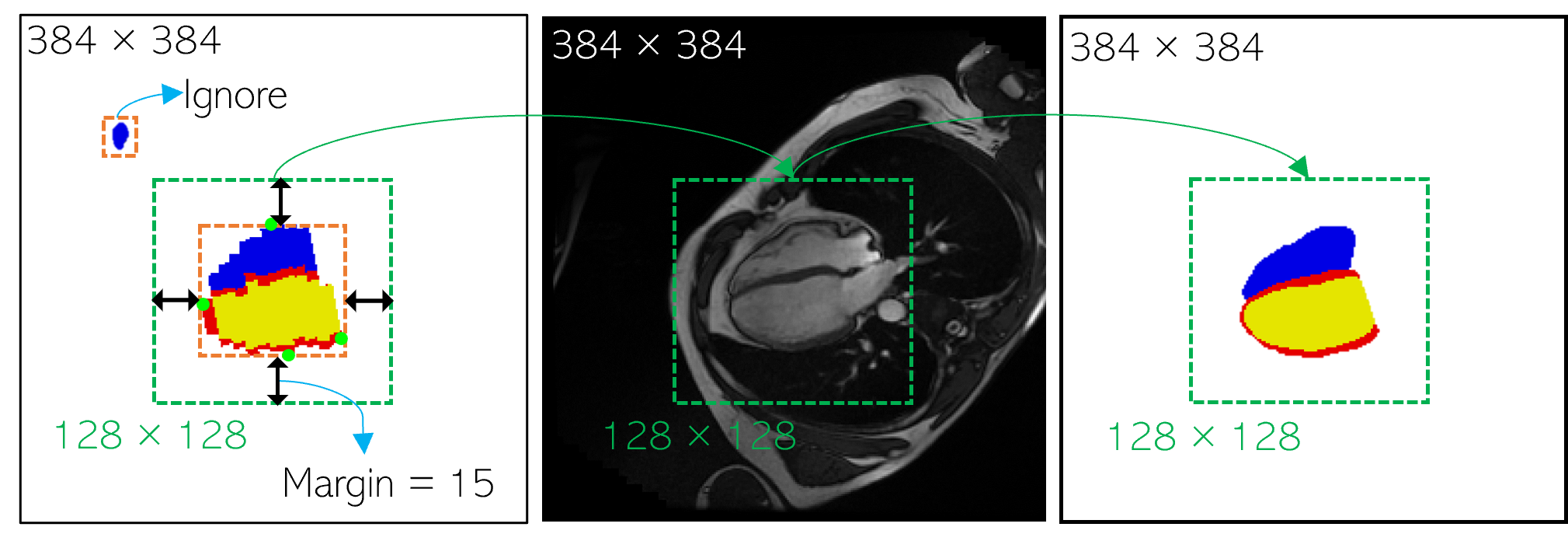}
\caption{Proposed Heart Localization and Cropping (HLC) module. The heart region is localized and cropped in both intensity and label images.}
\label{fig2}
\end{figure}
The segmentation networks utilizing full-scale images frequently confuse the background and cardiac tissue, resulting in considerable false positives. Using the proposed HLC module, the network only focuses on a smaller area containing heart tissues. Moreover, as the HLC module reduces the spatial resolution, the next stage network will run on a low-resolution image, leading to reduced computational complexity.\\
The HLC module is implemented by extracting the heart region containing the LV blood pool (LV), RV blood pool (RV), and left ventricular MYO from the original full-scale images. This can be achieved using the SA2LAmap for $I_{LA}$ and either LA2SAmap or $S_{SA1}$ for $I_{SA}$.


To find the bounding box across three labeled regions (LV-blood pool, RV-blood pool, and left ventricular MYO) within a segmentation prior, we'll use transformation maps (SA2LAmap, LA2SAmap, or $S_{SA1}$) as binary masks, where nonzero values represent the regions of interest. To ensure that the cropped region occupies the entire region and does not miss any pixels of the foreground regions, we defined a parameter named \emph{margin}, which can help to safeguard the edges and provide a margin that assists in preserving the entire heart in the cropped region, as shown in \figref{fig2}.

We further confirmed that the obtained cropped region from the HLC module safely encloses the entire heart by cropping the ground truth, restoring the original size, and finding the Dice score between the original size and the restored ground truth. We ensured a Dice score of 1 and Hausdorff Distance (HD) of 0 for the resized ground truth segmentation against the original size ground truth. Empirically, we found that a \emph{margin} of 15 pixels perfectly fits this purpose for all images in the training, validation, and test sets. We applied the same protocols to crop the intensity image and bring back the prediction from the cropped image to the original spatial dimension.

\subsection{The Transformation Process}
The M\&Ms-2 dataset is novel in terms of providing the images/labels of the LA view along with the SA view to give detailed information for the apical and basal slices of short-axis views \cite{martin2023deep}. We have utilized this information more efficiently in the proposed framework and generated the transformations for each axis. More specifically, we transformed one view's physical coordinates into the other view's image coordinates system and vice versa.

The trans-dimensional segmentaton prior \textbf{SA2LAmap} is obtained using the pseudo-code shown in Algorithm~\ref{algo1}. The prediction from TriggerNet $S_{SA1}$ has different metadata information than the original $I_{SA}$. This metadata information is essential for the conversion between physical coordinates and image coordinate systems and includes additional information like image orientation, voxel size, and origin. We used the CopyInformation function from SimpleITK to inherit all the relevant metadata from the original SA image to the $S_{SA1}$. This ensured that $S_{SA1}$ was 
\begin{algorithm}[H]
\textbf{Input}  $I_{LA}, I_{SA}, S_{SA1}, T_{SA \rightarrow LA}$ \\
\textbf{Output}  $SA2LAmap$
\begin{algorithmic}[]
\item[] Initialize output to zero, $SA2LAmap = 0$
\FOR{each point ($\mathbf{p}$) in $I_{LA}$}
\item Use $T_{SA \rightarrow LA}$ to transform $\mathbf{p}$ into $S_{SA1}$, producing $\mathbf{q}$
\item $SA2LAmap(\mathbf{p}) = S_{SA1} (\mathbf{q})$
\ENDFOR
\end{algorithmic}
\textbf{Return}  $SA2LAmap$
\caption{Transformation of the SA segmentation ($S_{SA1}$) to the viewpoint of the LA image ($I_{LA}$) using $T_{SA \rightarrow LA}$. } \label{algo1}
\end{algorithm}
\begin{algorithm}[H]
\textbf{Input}  $I_{LA}, I_{SA},S_{LA},T_{LA \rightarrow SA}$ \\
\textbf{Output}  $LA2SAmap$
\begin{algorithmic}[]
\item[] Initialize output to zero, $LA2SAmap = 0$
\FOR{each point $\mathbf{p}$ in $I_{SA}$}
\item Use $T_{LA \rightarrow SA}$ to transform $\mathbf{p}$ into $S_{LA}$, producing $\mathbf{q}$
\item $LA2SAmap(\mathbf{p}) = S_{LA} (\mathbf{q})$
\ENDFOR
\end{algorithmic}
\textbf{Return}  $LA2SAmap$
\caption{Transformation of the LA segmentation ($S_{LA}$) to the viewpoint of the SA image ($I_{SA}$) using $T_{LA \rightarrow SA}$. }\label{algo2}
\end{algorithm}
aligned correctly with the original $I_{SA}$ and could be used to transform the coordinate systems with the original $I_{LA}$.

The physical coordinates of the $I_{LA}$ are obtained from the image coordinate system, followed by finding the corresponding index in the $S_{SA1}$ and mapping these physical coordinates to the $S_{SA1}$ image coordinate system. Finally, it assigns the voxel value from the $S_{SA1}$ to the corresponding voxel in the SA2LAmap. In this way, the SA2LAmap is populated with all transformed values from $S_{SA1}$ to the SA2LAmap.

The trans-dimensional segmentation prior \textbf{LA2SAmap} is generated similarly, as shown in Algorithm~\ref{algo2}. In this case, the metadata information from the original $I_{LA}$ is copied to the $S_{LA}$ from LA-SegNet.
The physical coordinates of the $I_{SA}$ are extracted from its image coordinates system, followed by transforming this physical point to an index in the coordinate system of LA2SAmap.  Finally, each voxel value from the $S_{LA}$ is copied to the corresponding location in the LA2SAmap.

\subsection{Implementation Details}
The proposed architecture is implemented using a single NVidia A100 GPU with 40GB RAM. The SA and LA MRI scans are resampled to a voxel size of 1$\times$1$\times$1 ${mm^3}$. Dice loss and cross-entropy loss are used as loss functions to train the segmentation networks. Stochastic gradient descent is used as an optimizer with an initial learning rate of 0.01 and a Nesterov momentum of 0.99. We utilize a polynomial learning rate scheduler \cite{mishra2019polynomial} with a weight decay of 0.0005 to decrease the learning rate after each training epoch.
All networks, i.e., TriggerNet, LA-SegNet, SA-SegNet, and the nnUNet baseline, are trained independently for 1000 epochs (nnUNet default), where each epoch has 250 training iterations. The pre and post-processing steps, such as LA-2-SA transformation and vice versa, and the heart localization and cropping are performed in succession after the previous network predictions are available. All three segmentation networks and their respective pre and post-processing steps are carried out sequentially in the inference phase.  

Different data augmentation techniques are applied during training to allow the networks to see a stream of distinct examples. Spatial transformations, including random rotation, scaling, and mirroring, provide distinct spatial perspectives from which the model can learn. Intensity adjustments, such as random brightness, contrast, and gamma variations, ensure the model's adaptability to varying acquisition settings. Additionally, additive zero-mean Gaussian noise is utilized to enhance stochasticity, and blurring techniques, such as Gaussian blur, contribute to the model's robustness against variations in image quality.
\section{Dataset}
The Multi-Disease, Multi-View, and Multi-Center Right Ventricular Segmentation challenge (M\&Ms-2) was introduced in MICCAI 2021. The challenge focused on segmenting RV blood pools across cardiac imaging of multiple views and centers \cite{campello2021multi},\cite{martin2023deep}. The data includes diverse images from three clinical centers in Spain utilizing nine scanners from three vendors, including Siemens, General Electric, and Philips. It includes instances having various LV and RV pathologies as well as healthy subjects. The labels are provided for three regions of interest, including (i) LV blood pools, (ii) RV blood pools, and (iii) left ventricular MYO. It contains 360 instances from two cardiac cycles, specifically the end-diastolic and end-systolic phases. The subjects are divided sequentially into 160 for training, 40 for validation, and 160 for testing, such that different patients are in each split. The validation and test set also includes patients with pathologies not included in the training set. For each individual, both SA and LA MR images are provided, having SA and LA 4-chamber views.
\section{Ablation Studies} \label{ablation}
We study the effectiveness of our algorithmic design via different ablation studies. In particular, we evaluate the effect of utilizing the SA2LAmap and LA2SAmap/$S_{SA1}$ as a segmentation prior and the HLC module under different settings.

The SA2LAmap can be used in two ways to boost the network's performance: (i) \emph{Localization and Cropping Guide for HLC module:} To localize and crop the heart in original full-scale $I_{LA}$, and (ii) \emph{Segmentation Prior:} As a Segmentation prior concatenated to the $I_{LA}$. \tabref{tab1} lists the results of these experiments. 
\begin{table}[b!]
\centering
\caption{Ablation study to evaluate the effect of utilizing the Segmentation Prior and Heart Localization and Cropping (HLC) module. The first row also represents the baseline nnUNet results for LA.\\}\label{tab1}
        \resizebox{0.8\textwidth}{!}
        {
        \footnotesize
        \begin{tabular}{|c|c|c|c|c|c|c|c|} 
            \hline
              {HLC  Module} & \multirow{2}{*}{\pbox{3.5cm}{SA2LAmap As \\  Segmentation Prior}} &  \multicolumn{3}{|c|}{Dice Score LA  $\uparrow$}  & \multicolumn{3}{|c|}{HD (mm) LA $\downarrow$}\\[0.5ex] 
             \cline{3-8}
              & & LV & RV & MYO & LV & RV & MYO \\ 
            \hline
            \xmark & \xmark   & 0.94 & 0.90 & 0.86 & 5.91 & 6.61 & 5.98 \\
            \cmark & \xmark  & 0.95 & 0.91 & 0.87 & 3.81 & 4.90 & 3.08 \\
            \xmark & \cmark  & 0.95 & 0.92 & 0.86 & 3.26 & 4.35 & 2.73 \\
            \cmark & \cmark  & \textbf{0.96} & \textbf{0.93} & \textbf{0.88} & \textbf{2.81} & \textbf{3.80} & \textbf{2.51} \\
            \hline
        \end{tabular}
        }
\end{table}

\begin{figure}[b!]
\centering
\includegraphics[scale=0.65,trim={.2em, 0.5em, .0em, 0.0em}]{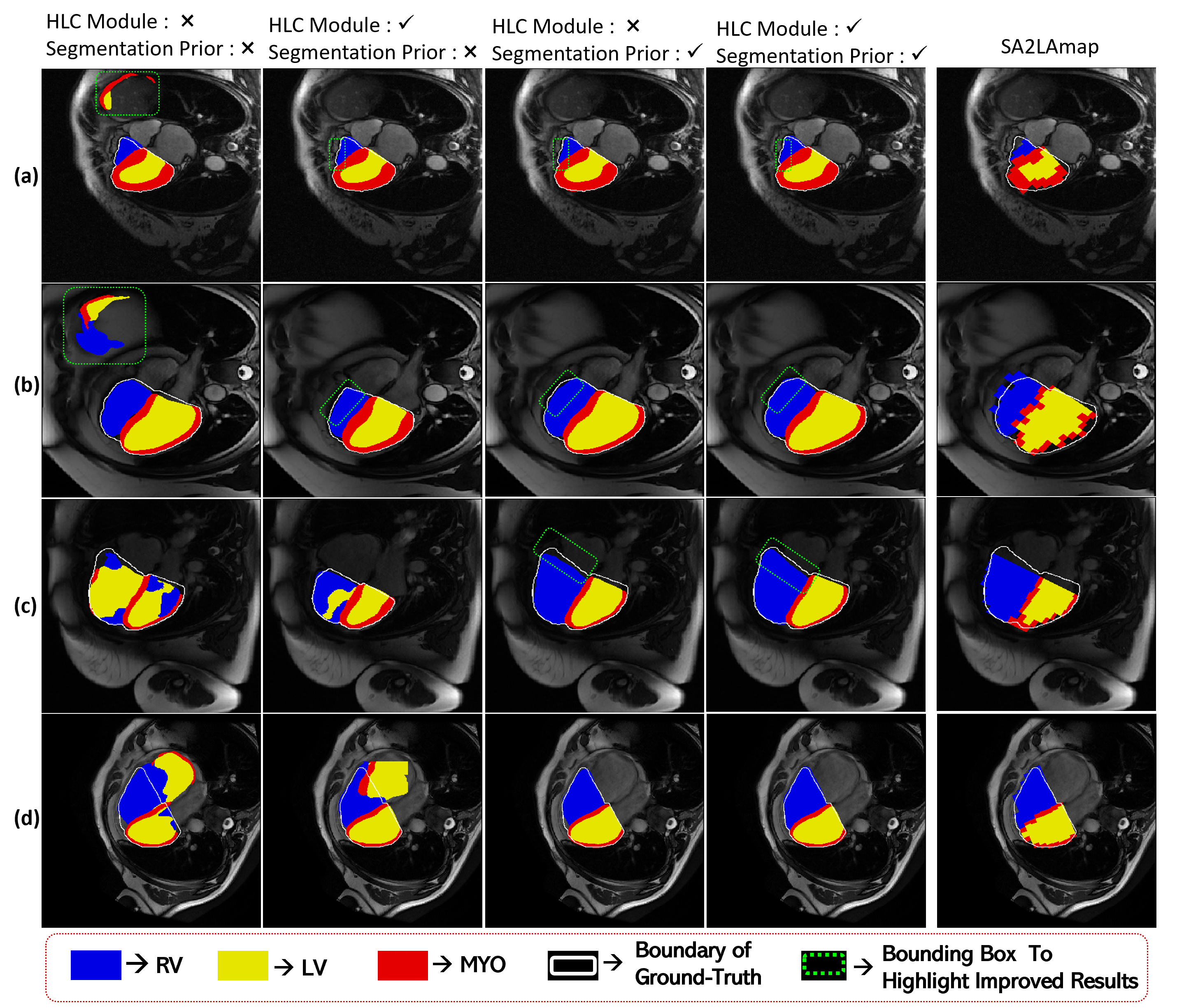}
\caption{Comparison of visual results under different settings for LA segmentation. From the left, the first column is the baseline results, and the second and third columns utilize HLC module and segmentation priors, respectively. The fourth column is the best results using both. The fifth column represents the SA2LAmap.
} 
\label{fig3}
\end{figure}
Row 1 represents the baseline where we do not use the SA2LAmap. Rows 2 and 3 show the results where we are using the SA2LAmap for the HLC module or as a segmentation prior, and row 4 depicts the performance where we first localized and cropped the heart in the $I_{LA}$ using the SA2LAmap and then concatenated the cropped image with the SA2LAmap. It can be seen that using either of the techniques (HLC module or segmentation prior) improves the results, which can be further validated with \figref{fig3}.
In \figref{fig3}, rows (a) and (b) depict that using the SA2LAmap either for the HLC module or as a segmentation prior can help remove the outliers (false positive predictions outside the area of interest). Rows (c) and (d) further confirm the usage of segmentation prior. The segmentation prior aids the network in generating anatomically plausible results. Thus, the SA2LAmap obtained from the $S_{SA1}$ helps to convert the invalid LA segmentation into close but correct shapes.
\begin{table}[t!]
\centering
\caption{Ablation study to evaluate the effect of utilizing the Segmentation Priors and Heart Localization 
 and Cropping (HLC) module. The first row also represents the baseline nnUNet results for SA.\\}\label{tab2}
        \resizebox{0.99\textwidth}{!}
        {
        \footnotesize
        \begin{tabular}{|c|c|c|c|c|c|c|c|c|} 
            \hline
             \multirow{2}{*}{\pbox{1.5cm}{HLC  Module}} & \multirow{2}{*}{\pbox{3.5cm}{$S_{SA1}$ As \\ Segmentation Prior}} & \multirow{2}{*}{\pbox{3.5cm}{LA2SAmap As\\ Segmentation Prior}} &  \multicolumn{3}{|c|}{Dice Score LA $\uparrow$}  & \multicolumn{3}{|c|}{HD (mm) LA $\downarrow$}\\[0.5ex] 
             \cline{4-9}
              & & & LV & RV & MYO & LV & RV & MYO \\ 
            \hline
            \xmark & \xmark  & \xmark  & 0.924 & 0.888 & 0.842 & 9.137 & 8.662 & 6.181 \\
            \cmark  & \cmark & \xmark  & 0.938 & 0.912 & \textbf{0.864} & 3.809 & 5.027 & 2.671 \\
            \cmark  & \cmark & \cmark  & \textbf{0.939} & \textbf{0.918} & 0.863 & \textbf{3.616} & \textbf{4.496} & \textbf{2.666} \\

            \hline
        \end{tabular}
        }
\end{table}

\tabref{tab2} showcases the ablation results for SA segmentation under various settings. Row 1 uses neither the HLC module nor the segmentation priors and lists the results of the TriggerNet. Row 2 lists the results of SA-SegNet if we only utilize the predictions from the TriggerNet to localize and crop the heart in the original full-scale image and use it as a concatenated form along with the image. For row 3, we also utilized the LA2SAmap. Using the LA2SAmap as segmentation prior improves all metrics except the MYO Dice score; here, we argue that the collective structure segmentation of MYO has already been improved in terms of structure variance by $S_{SA1}$ as segmentation prior, and LA2SAmap helps further eliminate errors in the boundary regions of MYO as indicated by the improved HD score.

\begin{figure}[b!]
\centering
\includegraphics[scale=0.75,trim={.2em, 0.5em, .0em, 0.0em}]{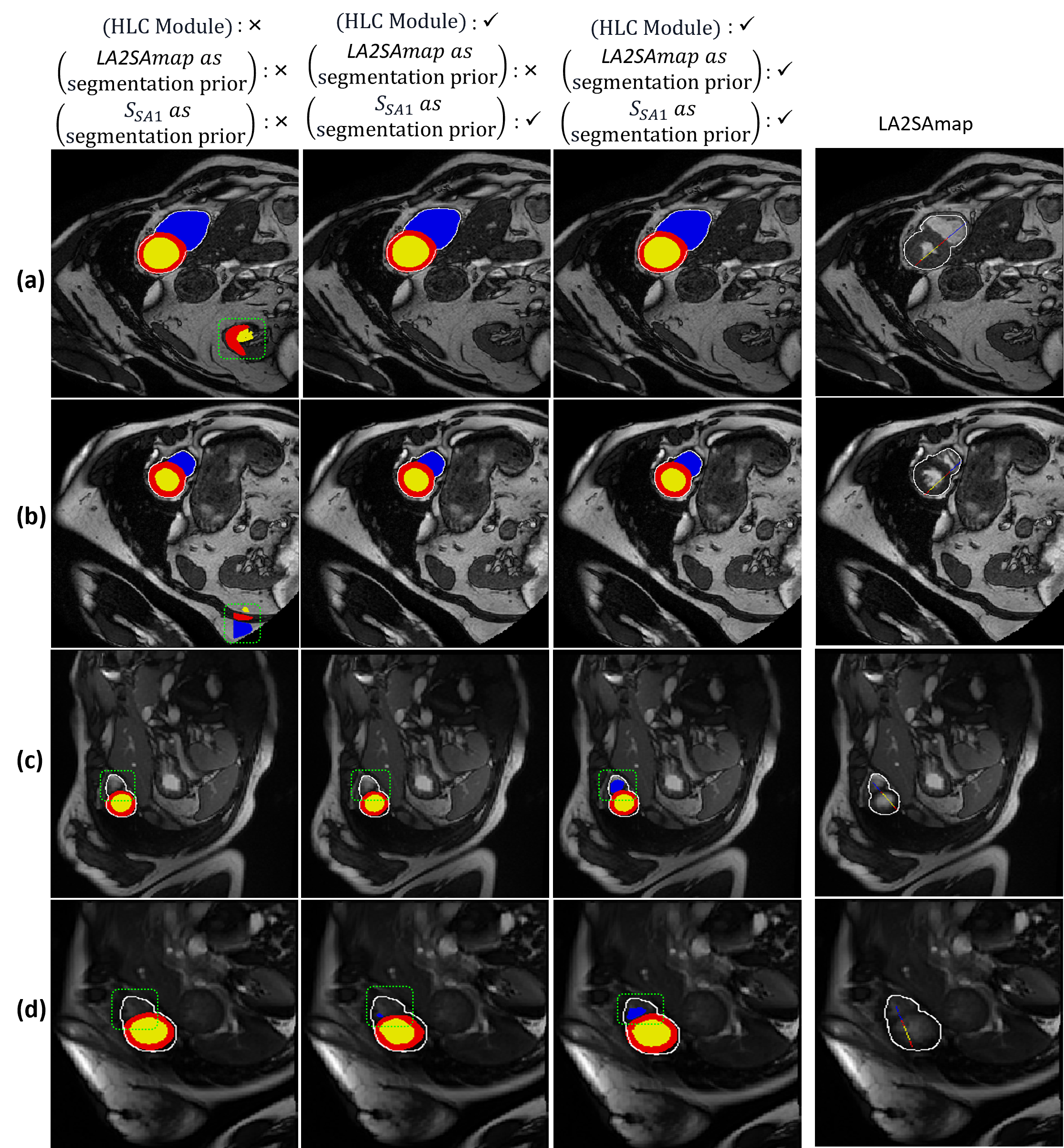}
\caption{Comparison of visual results under different settings for SA segmentation. From the left, the first column is the baseline results, and the second utilizes the HLC module and $S_{SA1}$ as segmentation priors. The third column shows the best results using both segmentation priors. The fourth column represents the LA2SAmap.} 
\label{fig4}
\end{figure}

\figref{fig4} depicts the visual results of this ablation study. Row (a) and (b) validate that each segmentation prior ($S_{SA1}$ or LA2SAmap) helps to remove the outliers. However, the last two rows, (c) and (d), further confirm that using the LA-to-SA transformation as a segmentation prior can further improve the segmentation. Here, we argue that in the cases where the TriggerNet cannot produce the accurate segmentation map for any of the regions of interest, then LA-SegNet might have already segmented that instance accurately with valid anatomical shape prior, and we can transfer that knowledge to LA2SAmap and is used it as segmentation prior for the SA-SegNet.\\
\figref{param} further advocates that the HLC module not only improves the segmentation performance but also decreases the overall computational requirements of the proposed pipeline. In \figref{param}, all the networks using the HLC module to localize the heart region have fewer encoder-decoder stages than those utilizing full-scale images, resulting in fewer parameters and Giga Multiply-Accumulate Operations (GMac). This is because if the in-plane spatial resolution size for input images is 384 $\times$ 384, to get bottleneck features of in-plane spatial dimension = 6 $\times$ 6, the encoder needs to compress the features six times; however, if the heart is cropped and localized using HLC module, it will produce the images of in-plane spatial dimension = 192 $\times$ 192, and the encoder will perform the feature compression in five instances to obtain bottleneck features with in-plane spatial dimension of 6  $\times$ 6.

\section{Results and Discussion}
We compared the results of the proposed approach with several state-of-the-art architectures using a five-fold cross-validation method as well as comparing against the challenge leaderboard using the provided validation and test set.

\begin{figure}[t!]
\centering
\includegraphics[scale=0.6,trim={.2em, 0.5em, .0em, 0.0em}]{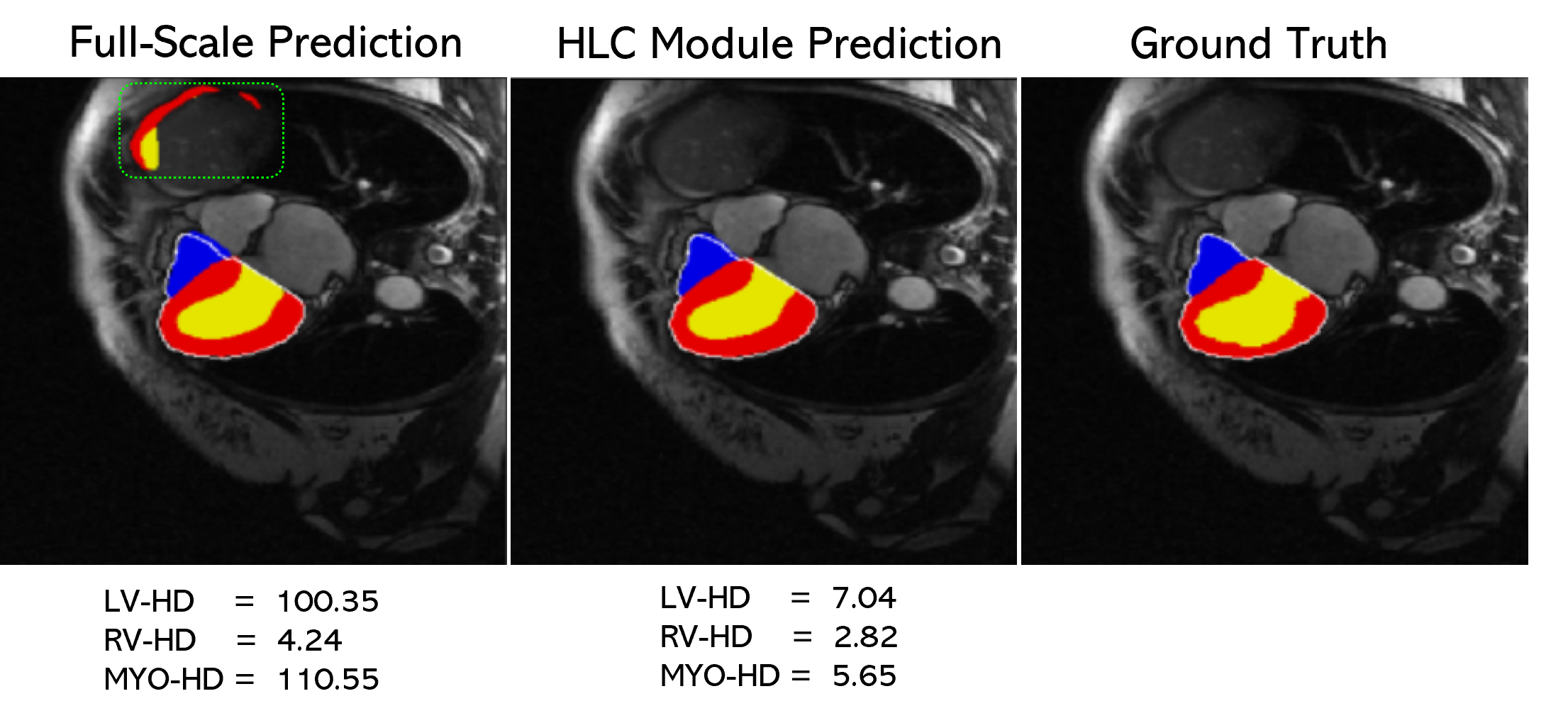}
\caption{Visual and quantitative analysis of how the outliers contribute to the higher HD, and comparison of results obtained from methods utilizing the full-scale images vs the proposed HLC module.} 
\label{fig5}
\end{figure}

\begin{table*}[t!]
\caption{Comparison to state-of-the-art methods for long/short-axis MRI on five-fold cross-validation split on the M\&MS-2 challenge dataset. The best and second results are in \textbf{bold} and \underline{underline}, respectively. Methods indicated with $*$ use multi-view inputs.\\}\label{tab3}
        \resizebox{1.0\textwidth}{!}{
        \footnotesize
        \begin{tabular}{|c|c|c|c|c|c|c|c|c|}
            \hline
             \multirow{2}{*}{Methods} &  \multicolumn{4}{|c|}{Dice($\%$)-Short-axis $\uparrow$}  & \multicolumn{4}{|c|}{HD (mm)-Short-axis $\downarrow$} \\
             \cline{2-9}
              & LV & RV & MYO & Avg & LV & RV & MYO & Avg \\ 
            \hline
            UNet   & 87.02 & 88.85 & 79.07 & 84.98 & 13.78 & 12.10 & 12.23 & 12.70 \\
            ResUNet   & 87.98 & 89.63 & 79.28 & 85.63 & 13.80 & 11.61 & 12.09 & 12.50 \\
            DLA   & 87.27 & 89.88 & 80.23 & 86.12 & 13.25 & 10.84 & 12.31 & 12.13 \\
            InfoTrans*   & 88.24 & 90.41 & 80.25 & 86.30 & 12.41 & 10.98 & 12.83 & 12.07 \\
            rDLA*   & 88.64 & 90.28 & 80.78 & 86.57 & 12.74 & 10.31 & 12.49 & 11.85 \\
            TransUNet   & 87.91 & 88.69 & 78.67 & 85.09 & 13.80 & 10.29 & 13.43 & 12.51 \\
            MCTrans   & 88.52 & 89.90 & 80.08 & 86.17 & 12.29 & 9.92 & 13.28 & 11.83  \\
            MCTrans*   & 87.79 & 89.22 & 79.37 & 85.46 & 11.28 & 9.39 & 13.84 & 11.49 \\
            UTNet   & 87.52 & 90.57 & 80.20 & 86.10 & 12.03 & 9.78 & 13.72 & 11.84 \\
            UTNet*   & 87.74 & 90.82 & 80.71 & 86.42 & 11.79 & \underline{9.11} & 13.41 & 11.44 \\
            TransFusion*   & \underline{89.52} & \underline{91.75} & \underline{81.46} & \underline{87.58} & \underline{11.31} & 9.18 & \underline{11.96} & \underline{10.82} \\
            \textbf{Proposed*}   & \textbf{93.94} & \textbf{91.87} & \textbf{86.34} & \textbf{91.71} & \textbf{3.61} & \textbf{4.49} & \textbf{2.66} & \textbf{3.58} \\
            \hline

              &  \multicolumn{4}{|c|}{Dice($\%$)-Long-axis $\uparrow$} & \multicolumn{4}{|c|}{HD (mm)-Long-axis $\downarrow$}\\
             \cline{2-9}
              &  LV & RV & MYO & Avg & LV & RV & MYO & Avg\\ 
            \hline
            UNet   & 87.26 & 88.20 & 79.96 & 85.14 & 13.04 & 8.76 & 12.24 & 11.35\\
            ResUNet   &  87.61 & 88.41 & 80.12 & 85.38 & 12.72 & 8.39 & 11.28 & 10.80\\
            DLA   &  88.37 & 89.38 & 80.35 & 86.03 & 11.74 & 7.04 & 10.79 & 9.86\\
            InfoTrans*   &  88.21 & 89.11 & 80.55 & 85.96 & 12.47 & 7.23 & 10.21 & 9.97\\
            rDLA*   &  88.71 & 89.71 & 81.05 & 86.49 & 11.12 & 6.83 & 10.42 & 9.46\\
            TransUNet   & 87.91 & 88.23 & 79.05 & 85.06 & 12.02 & 8.14 & 11.21 & 10.46\\
            MCTrans   &  88.42 & 88.19 & 79.47 & 85.36 & 11.78 & 7.65 & 10.76 & 10.06\\
            MCTrans*   & 88.81 & 88.61 & 79.94 & 85.79 & 11.52 & 7.02 & 10.07 & 9.54\\
            UTNet   & 86.93 & 89.07 & 80.48 & 85.49 & 11.47 & 6.35 & 10.02 & 9.28\\
            UTNet*   &  87.36 & 90.42 & 81.02 & 86.27 & 11.13 & 5.91 & 9.81 & 8.95\\
            TransFusion*   &  \underline{89.78} & \underline{91.52} & \underline{81.79} & \underline{87.70} & \underline{10.25} & \underline{5.12} & \underline{8.69} & \underline{8.02}\\
            \textbf{Proposed*}  & \textbf{95.80} & \textbf{93.07} & \textbf{87.71} & \textbf{92.19} & \textbf{2.81} &\textbf{3.80} & \textbf{2.51} & \textbf{3.04}\\
            \hline
            
        \end{tabular}
        }
\end{table*}

\begin{table*}[h!]
\caption{Performance comparison with the M\&Ms-2 challenge leaderboard for RV segmentation.\\}\label{tab4}
\centering
\resizebox{1.0\textwidth}{!}{
\begin{tabular}{|c|c|c|c|c|}
    \hline
     \multicolumn{5}{|c|}{\textbf{Quantitative comparison on validation set}} \\
    \hline
     Methods &  Dice Score LA $\uparrow$ & HD-LA(mm) $\downarrow$ &  Dice Score SA $\uparrow$ & HD-SA(mm) $\downarrow$ \\
    \hline
    \cite{sun2022right}   & {0.922} & 5.35 & {0.925} & 8.90 \\
    \cite{arega2022using}   & 0.922 & 5.59 & 0.924 & 8.85 \\
    \cite{li2021right}   & 0.920 & 5.34 & 0.922 & 9.47 \\
    
    Proposed   & \textbf{0.926} & \textbf{3.49} & \textbf{0.928} & \textbf{3.72} \\
    \hline
    \multicolumn{5}{|c|}{\textbf{Quantitative comparison on test set}}\\
    \hline
     Methods & Dice Score LA $\uparrow$ & HD-LA(mm) $\downarrow$ &  Dice Score SA $\uparrow$ & HD-SA(mm) $\downarrow$ \\
    \hline
    \cite{sun2022right}   & {0.919} & 6.04 & {0.925} & 10.58 \\
    \cite{arega2022using}   & 0.919 & 6.10 & 0.920 & {9.94} \\
    \cite{li2021right}   &  0.916 & 6.17 & 0.920 & 10.30 \\
    
    Proposed   & \textbf{0.928} & \textbf{3.91} & \textbf{0.927} & \textbf{4.01} \\
    \hline
\end{tabular}
}
\end{table*} 

\begin{table*}[t!]
\caption{Performance comparison with multi-stage segmentation approach \cite{al2023reducing} evaluated on 200 subjects of validation and test set.\\}
\centering
\label{tab5}
        \resizebox{0.8\textwidth}{!}{
        \footnotesize
        \begin{tabular}{|c|c|c|c|c|c|c|}
            \hline
             \multirow{2}{*}{Methods} &  \multicolumn{3}{|c|}{Dice($\%$)-Short-axis $\uparrow$}  & \multicolumn{3}{|c|}{HD (mm)-Short-axis $\downarrow$} \\
             \cline{2-7}
              & LV & RV & MYO  & LV & RV & MYO  \\ 
            \hline
            \cite{al2023reducing}    & {0.959} & \textbf{0.938} & \textbf{0.907}  & {6.42} & 8.62 & {9.37} \\
            {Proposed}   & \textbf{0.963} & {0.928} & {0.870} & \textbf{3.43} & \textbf{3.87} & \textbf{3.62}  \\
            \hline

              &  \multicolumn{3}{|c|}{Dice($\%$)-Long-axis $\uparrow$} & \multicolumn{3}{|c|}{HD (mm)-Long-axis $\downarrow$}\\
             \cline{2-7}
              &  LV & RV & MYO  & LV & RV & MYO \\ 
            \hline
            \cite{al2023reducing}   & {0.958} & {0.924} & \textbf{0.901} & {4.07} & 5.81 & {5.27} \\
            {Proposed}   & \textbf{0.961} & \textbf{0.927} & {0.878}  &\textbf{2.83}  & \textbf{3.70} & \textbf{2.56} \\
            \hline
            
        \end{tabular}
        }
\end{table*} 

\tabref{tab3} lists the comparison of the results with other segmentation architectures in terms of the Dice score and HD on a five-fold cross-validation split. The proposed method is able to produce robust segmentation across both SA and LA views. A substantial improvement can be seen in terms of reducing the HD score. This gain in performance comes from the fact that the proposed HLC module can remove the outliers efficiently, resulting in an accurate segmentation map compared to the full-scale image methods. \figref{fig5} further illustrates this effect.
Compared to full-scale prediction, the segmentation map generated by utilizing the proposed HLC module eliminates all the erroneous predictions outside the region of interest. We can further observe the variation in HD score for the predictions with and without the outliers. It can be seen that the difference in HD score for MYO and LV is very large compared to the RV because of the false positives.

\tabref{tab4} provides results compared to the challenge leaderboard on the validation set and the test set subjects for RV segmentation. This further proves the generalization ability of the proposed approach to the unseen pathologies, as the validation and test set has two pathologies (Tricuspidal Regurgitation and Congenital Arrhythmogenesis) not available in the training set.

We also compared the proposed approach with other multi-stage segmentation methods, such as those mentioned in \cite{al2023reducing}, and \tabref{tab5} lists the comparison results. Similar to \cite{al2023reducing}, the comparison is performed on the 200 subjects by combining the validation (40) and testing (160) examples, while the algorithm's development occurs solely based on the training data. The proposed method outperforms \cite{al2023reducing} for most of the evaluation metrics.

\subsection{Limitations and future work}
The proposed pipeline is useful in the settings where we have cardiac MRI available for both LA and SA views. Clinically, both views are captured according to cardiac MRI acquisition protocol \cite{kramer2020standardized}, \cite{petersen2015uk}; however, most of the publicly available datasets only provide short-axis cine MR images and their labels \cite{campello2021multi}. This unavailability of long-axis MR images can be considered a limitation of the proposed approach for datasets with single-view images only. However, we expect future cardiac MRI datasets to release more complementary information, such as both views, to take advantage of their relationship. We also encourage the research community to provide a 2-chamber and 3-chamber LA view, further exploiting the multi-view aspect of cardiac MR images.

Finally, we analyze that a medical image is influenced by the anatomical features of the image and the characteristics of the imaging equipment, such as vendor information, scanner type, etc. This inspires our future work, where we will aim to design more robust pipelines to incorporate metadata along with intensity images for segmentation tasks \cite{lemay2021benefits}. This will enable the segmentation networks to learn not only the appearance of images but also the specific interdependence of an image structure and image-capturing device.

\section{Conclusion}
This paper proposes a cardiac image segmentation approach relying on the trans-dimensional segmentation priors between short-axis and long-axis views. We show that the method provides a substantial improvement in the accuracy of segmentation for cardiac images in LV, RV blood pools, and left ventricular MYO. The proposed approach effectively utilizes the relationship between the SA and LA views so that a segmentation in one view informs the segmentation in the other view. The transformed maps are used to localize and crop the heart region in the original full-scale image using the HLC module of the same axis and act as a segmentation prior to the other axis. The HLC module helps to remove the outliers and improves erroneous predictions. The segmentation prior encourages anatomically plausible segmentation maps. Extensive ablation studies are conducted to show the efficacy of proposed techniques, and the results are compared with the existing state-of-the-art methods utilizing the M\&Ms-2 dataset.

\section*{Acknowledgments}
This work was funded through the mini-Centre for Doctoral Training in AI-based Cardiac Image Computing provided through the Faculty of Science and Engineering, Queen Mary University of London.  This paper utilised Queen Mary's Andrena HPC facility. Caroline Roney acknowledges funding from a UKRI Future Leaders Fellowship (MR/W004720/1). 
This work also acknowledges the support of the National Institute for Health and Care Research Barts Biomedical Research Centre (NIHR203330), a delivery partnership of Barts Health NHS Trust, Queen Mary University of London, St George’s University Hospitals NHS Foundation Trust and St George’s University of London.

\bibliographystyle{elsarticle-num} 
\bibliography{references}





\end{document}